\documentstyle[12pt,aasms]{article}
\begin{document}
\title{Microlensing by the Galactic Bar}
\author{HongSheng Zhao,}
\affil{Dept. of Astronomy, Columbia University, NY, NY 10027 USA}
\author{David N. Spergel,}
\affil{Princeton University Observatory, Princeton, NJ 08544 USA}
\centerline{and}
\author{R. Michael Rich}
\affil{Dept. of Astronomy, Columbia University, NY, NY 10027 USA}
\begin{abstract}
	We compute the predicted optical depth and duration
distribution of microlensing events towards Baade's window in a model
composed of a Galactic disk and a bar.  The bar model is a
self-consistent dynamical model built out of individual orbits that has been
populated to be consistent with the {\sl COBE} maps of the Galaxy
and kinematic observations of the Bulge.  We find that most of the lenses
are in the Bulge with a line-of-sight distance 6.25 kpc (adopting
$R_0=8$ kpc).  The microlensing optical depth of a $2\times
10^{10}M_\odot$ bar plus a truncated disk is $(2.2\pm 0.3)
\times 10^{-6}$, consistent with the very large optical depth
$(3.2\pm 1.2) \times 10^{-6}$ found by Udalski et al. (1994).  This
model optical depth is enhanced over the predictions of axisymmetric
models by Kiraga \& Paczy\'nski (1994, hereafter KP) by slightly more
than a factor of two since the bar is elongated along the
line-of-sight.  The large Einstein radius and small transverse
velocity dispersion also predict a longer event duration in the
self-consistent bar model than the KP model.  The event rate and
duration distribution also depend on the lower mass cutoff of the
lens mass function.  With a $0.1M_\odot$ cutoff, 5-7 events (depending
on the contribution of disk lenses) with a logarithmic mean duration
of 20 days are expected for the OGLE experiment according to our
model, while Udalski et al. (1994) observed 9 events with durations
from 8 to 62 days.  On the other hand, if most of the lenses are brown
dwarfs, our model predicts too many short duration events.  A KS test
finds only 7\% probability for the model with $0.01M_\odot$ cutoff
to be consistent with current data.
\end{abstract}
\vfill
\eject

	Microlensing experiments were proposed to solve one of the
outstanding problems of astrophysics: the dark matter problem.  They
aim to detect the massive compact halo objects (MACHOs) that have been
suggested as the dominant mass component of our Galaxy (Paczy\'nski
1986; Griest et. al. 1991) through their microlensing of distant stars
and the composition of the disk through observations of the bulge
(Paczy\'nski 1991).  Like many astronomical observations, these
experiments appear to raise new questions rather than solve
outstanding problems.  For while experiments that look for the halo
dark matter by observing stars in the LMC appear to report too low an
event rate for MACHOs to be the halo dark matter (Gould 1994),
experiments that monitor stars in the Galactic Bulge appear to detect
too high an event rate (Udalski et al.  1994).

	Both the OGLE and the MACHO microlensing experiments find an
unexpectedly large number of microlensing events towards the Galactic
Bulge: 34 at the time of submission.  Udalski et al. (1994) have
analyzed the OGLE data and derived a very large lensing optical depth
$\tau = (3.3 \pm 1.2) \times 10^{-6}$ towards the OGLE fields: Baade's
window ($l = 1^o, b = -4^o$) and two adjacent fields ($\pm 5^o$,
$-4^o$).  KP and Guidice et al. (1994) find that lensing of bulge
stars by stars in the disk can account for at most 20\% of this
observed optical depth.  When KP include an axisymmetric bulge in
their lensing calculations, they find that it is the dominant source
of lens events.  However, even their bulge plus disk model can account
for only $\sim 30\%$ of the observed events and it predicts too short
an event duration.  KP suggest that this discrepancy may be due to
their modeling the bulge as an axisymmetric rotator (Kent 1992),
rather than as a bar.  However, they did not make quantitative
calculations of the prediction of bar models.

	There is a growing consensus in the astronomical community
that the Milky Way is a barred Galaxy.
  Binney et al. (1991) argue that a bar could explain the non-circular
motions seen in both the CO and HI observations.  Star counts (Nakada
et al.  1992, Whitelock and Catchpole 1992, Stanek et al. 1994) find
that the characteristic magnitudes of bulge stars at positive
longitudes are larger than bulge stars at negative longitudes,
consistent with the bar hypothesis.  Blitz and Spergel (1991) suggest
that the asymmetries between the first and fourth quadrants in the IR
surface brightness distribution from the balloon observation of
Matsumoto et al. (1982) implies that the Galaxy is barred.  These
asymmetries are confirmed by recent DIRBE multicolor maps of the
Galaxy (Weiland et al. 1994).  Dwek et al.  (1994) use these DIRBE
maps to construct a three-dimensional triaxial model of the bulge,
which we use to compute the optical depth for lensing towards Baade's
window.

	As part of the doctoral thesis, Zhao (1994) has developed a
self-consistent model for the Galactic bar following Schwarzschild's
(1979) method.  The density profile in this model follows the Dwek et
al. (1994) fit to the {\sl COBE} image of the galaxy.  This model is
constructed by running 6000 stellar orbits for $\sim 3.5$ Gigayears
and then weighting these stellar orbits to match observations.  In
this {\sl Letter }, we use this state-of-the-art bulge model to
compute the predicted optical depth and event duration for
microlensing towards the bulge.

	According to KP, the optical depth averaged over all
detectable stars is computed from
\begin{equation}
\tau = {1 \over \int_0^\infty ds \, w(s)}
                \int_0^\infty ds \, w(s) \int_0^s dl \rho(l) D \, ,
\end{equation}
where $w(s)$ is the probability of the source being at a distance $s$,
$\rho(l)$ is the lens mass density at a distance $l$, and $D \equiv
(s-l)l s^{-1}$ is the characteristic distance between the lens and
source.  The averaging over the source distance distribution is
necessary, because there are events that both the lens and the source
are in the bulge due to the finite depth of the bulge.  Following KP,
we adopt a power law luminosity function with the fraction of stars
more luminous than $L$ being proportional to $L^{-\beta}$.  For a
magnitude limited survey, this implies that $w(s) = \rho(s) s^{2
-2\beta}$; this is probably valid within the source distance range 4
to 12 kpc.  We derive a raw luminosity function from the
Color-Magnitude data of Paczy\'nski
et al. (1994), and find a good fit when $\beta$ is between 0.75 and 1.
Also in the range of luminosities near the magnitude limit of the
OGLE, our fit to the Terndrup et al. (1990)
star counts implies $\beta \approx 1.5 $. Let us set $\beta=1\pm 0.5$;
a smaller $\beta$ would make a slightly larger optical depth.

	Dwek et al. (1994) has fit a series of luminosity density
models to the DIRBE surface brightness observations at $|b| > 3^o$.
Their best fit model are the triaxial Gaussian-type models (the G1 and
G2 models).  The G2 model is boxy and has a density
\begin{equation}
	\rho(x,y,z) = {M \over 8 \pi a b c } \exp\left(-{s^2 \over 2} \right) \, ,
\end{equation}
where
\[
s^4 =\left[\left({x \over a}\right)^2 + \left({y \over b}\right)^2
\right]^2 +  \left({z \over c}\right)^4 \, .
\]
The scale lengthes $a = 1.49 \pm 0.05$, $b = 0.58 \pm 0.01$ and
$c=0.40 \pm 0.01$ kpc for galactocentric distance $R_0$=8 kpc.  The long
axis of the bar, the $x$ axis, points towards $l = -13.4^o$.

The total luminosity in Dwek et al.'s G2 model, as well as
observations of stellar and gas kinematics (Kent 1992) imply a bulge mass
of $\sim 1 - 2 \times 10^{10} M_\odot$.  Using this, we estimate
the optical depth of the bar at Baade's window to be
\begin{equation}
\tau_{bar} = (0.84 \pm 0.1) \times 10^{-6} M_{10} \, ,
\end{equation}
where $M_{10}$ is the bulge mass in units of $10^{10} M_\odot$.  When
using their ellipsoidal Gaussian model (G1 model) instead, we find a 5\%
reduction in optical depth.  $\tau_{bar}$ is also 25\% smaller for the
two adjacent fields ($l=\pm 5^o$, b=$-4^o$).

The halo is not expected to make a significant contribution to the
optical depth towards Baade's window.  Griest (1991) estimates the
halo's optical depth towards the Bulge $\tau_{halo} \leq 0.14 \times
10^{-6}$ for a reasonable halo core radius $a \geq 4$ kpc.
Also, the events due to halo objects are likely to be too short
to be detectable by OGLE (Udalski et al. 1994).

The disk, however, is expected to make a significant contribution to
the optical depth towards Baade's window.  KP note that in the
Bahcall-Soneira (BS) model the disk density along the line-of-sight
towards $b = -4^o$ is nearly constant.  On the other hand, infrared
observations of the Galaxy (Kent, Dame, \& Fazio 1991) suggest a
smaller disk scale length than 3.5 kpc in the BS model; this would
imply that the stellar density increases along Baade's window
line-of-sight.  We compute the optical depth for a double exponential
disk normalized locally by the disk stellar density $\sim 0.1 {\rm
M_\odot pc}^{-3}$ (Bahcall, Flynn and Gould 1992) and the surface
density of $71\pm 6{\rm M}_\odot {\rm pc}^{-2}$ (Kuijken and Gilmore
1991).  We allow the disk to be truncated at some distance
(Paczy\'nski et al. 1994).  For disk models with the full range of
reasonable scale length from 2.7 kpc to 3.5 kpc (Kent, Dame, \& Fazio
1991, Bahcall and Soneira 1980), the optical depth $\tau_{disk}$ is
from $0.87 \times 10^{-6}$ to $0.63 \times 10^{-6}$ for the full disk
model, and from $ 0.47 \times 10^{-6}$ to $ 0.37 \times 10^{-6}$ for a
disk truncated at 4 kpc.

If we sum the contributions of a $2\times 10^{10}{\rm M}_\odot$ bar
and a truncated disk with 2.7 kpc scale length ($\tau_{disk}=0.47
\times 10^{-6}$), then the predicted optical depth $(2.2 \pm 0.3)
\times 10^{-6} $ lies within the error range of the optical depth
determined by Udalski et al. (1994) from the OGLE data: $(3.3 \pm 1.2)
\times 10^{-6}.$  The error bar in the observed optical depth will be
significantly reduced when the detection efficiency of MACHO
experiment is quantified.

	Unlike the optical depth, the event time scale depends on both
the lens mass function and the velocity distribution of lens and
source.  KP note that in their axisymmetric bulge model, most of the
events are expected to be of such short duration that they would not
be detected in the OGLE experiment.  Here, we compute the predicted
event distribution in our bar plus truncated disk model and determine
whether the predicted events are consistent with the OGLE
observations.

	Following KP, we adopt a logarithmic mass function
between a mass range of $10^{-\gamma} M_\odot \leq m \leq M_\odot$
and treat $\gamma$ as a free parameter.    The differential
lensing duration distribution is then determined by the density distribution
and the phase space distribution,
\begin{equation}
P(t_0) \equiv {d\Gamma(t_0) \over d\log(t_0)} = {16 G \epsilon(t_0)\over \gamma
c^2 t_0}
<\int_0^s g(v,D) \rho(l) D dl>
\end{equation}
Here the average is over the source distance.  $G$ and $c$ are the
gravitational constant and the speed of light.  $\epsilon(t_0)$ is the
observation detection efficiency of events of time scale $t_0 \equiv
R_E/v$; for OGLE, we find $ \epsilon(t_0) = 0.3 \exp (-(t_0/11{\rm
day})^{-0.7}) $ is a convenient and good interpolation of values given
in Udalski et al. (1994).  The dimensionless factor $g$ is the phase
space fraction of sources and lenses whose relative proper motion
velocity satisfies
\begin{equation}
2 r_{\rm low} D \leq v^2 t_0^2 \leq 2 r_{\rm upp} D \, ,
\end{equation}
where $r_{\rm low}= 2GM_{\rm low}c^{-2}$ and $r_{\rm upp} = 2 GM_{\rm
upp}c^{-2}$ are the Schwarzchild radius corresponding to the lower and
upper mass cutoffs.  In our calculations, we evaluate $g$ by
Monte-Carlo integration over the six-dimensional phase space.

Before turning to our self-consistent bar model to compute $g$, we can
estimate the event time distribution by approximating the relative
proper motion distribution as a 2-D Gaussian of transverse velocity
dispersion $\sigma_t$:
\begin{equation}
g \approx \exp\left(-{r_{\rm low} D \over \sigma^2 t_0^2}\right)
-\exp\left(-{r_{\rm upp} D \over \sigma^2 t_0^2} \right)
\approx \exp\left(-{r_{\rm low} D \over \sigma^2 t_0^2}\right) \, .
\end{equation}
The steep drop-off of $g$ and $\epsilon(t_0)$ for short events, together with
the $t_0^{-1}$
drop-off for long events, implies that $P(t_0)$ peaks near
\begin{equation}
t_p = (t^2_\epsilon+ R_E^2\sigma_t^{-2} )^{0.5} \, ,
\end{equation}
where the time scale $t_\epsilon$= 7 days is due to the steep drop-off
of OGLE detection efficiency for very short events, and the
characteristic Einstein radius $R_E$ = (80 day $\times$ 100
km s$^{-1}) M_{\rm low}^{0.5} D^{0.5}_{\rm kpc}$.  A more massive lens and a
larger distance between the lens and the source makes a larger
Einstein radius.  Together with a lower transverse dispersion, it
shifts the distribution towards longer duration.  The Spaenhauer et
al.  (1992) analysis of proper motion data in Baade's window finds
$\sigma_t \equiv (\sigma^2_l+\sigma^2_b)^{1/2} = 150$ km s$^{-1}$.  We
estimate that the average characteristic distance $D\simeq 0.75$ kpc
in the Dwek G2 model.  So a model with most lenses being brown dwarfs
and the lower mass cutoff at $10^{-2} M_\odot$ would predict a peak in
duration distribution at about 8 days, while OGLE detected events with
duration ranging from 8 to 62 days.

We can improve our estimate of event distribution by using Zhao's
(1994) self-consistent bar which fits the G2 model of Dwek et al.
(1994), the radial velocity and proper motion dispersions at Baade's
window (Sharples et al. 1990, Spaenhauer et al. 1992, Zhao et al.
1994) and a mean stellar rotation curve of slope 60 km
s$^{-1}$kpc$^{-1}$ (e.g., Izumiura et al. 1992).  Our galactic
potential consists of the G2 model for the bar, a Miyamoto-Nagai
potential for the disk and an isothermal dark halo (Binney \& Tremaine
1987).  The bar mass is fixed at $2 \times 10^{10} M_\odot$ with a
pattern speed of 60 km s$^{-1}$kpc$^{-1}$ similar to the Binney et
al. (1991) model.  The disk parameters are fit to the BS model and the
halo parameters are fixed so that the rotation curve is flat out to 20
kpc.  The stellar distribution function is composed of the weighted
sum of 6000 orbits, each of which has been run for 1024 orbit
crossings.  Quadratic programming (e.g., Merritt 1993) is used to
assign weights to each of these orbits so that their sum reproduces
the G2 model and the observed kinematics.

In Figure 1, we plot the transverse velocity dispersion $\sigma_t$
(solid line) and the absolute value of transverse rotation speed $V_t$
(dashed line) in km s$^{-1}$ from Zhao (1994) bar model, and the
arbitrarily scaled probability of lens location (dotted line) as
functions of distance from the Sun along Baade's window line-of-sight.
The probability includes the disk lenses.  As both the disk and the
bulge are truncated at 4 kpc, there is a break in lens density at the
truncation point.  Note that most of the lenses are at 6.25 kpc in the
Bulge, well in front of the sources.  The large Einstein radius,
together with the low transverse velocity dispersion at the most
probable location of lenses, shifts the event distribution towards
longer duration in the Galactic bar model.

Using this bar model, we can directly evaluate $g$ by a Monte-Carlo
integration over the stellar distribution function.  Combining this
result with the reported OGLE efficiencies yields our prediction for
the event duration distribution in the OGLE experiment.  A truncated
small scale length disk is also included in the calculation with a
disk velocity distribution same as KP.  Figure 2 shows event duration
distributions of the OGLE data (histogram) from Udalski et al. (1994)
and models with the lower mass cutoffs at $10^{-1} M_\odot$ (solid
line) and $10^{-2} M_\odot$ (dotted line).  The upper panel shows the
normalized cumulative distribution.  The lower panel shows the
differential distribution.  The model with cutoff at $10^{-2} M_\odot$
predicts too many short duration events; a KS test finds only a 7\%
probability that it is consistent with the data.  The observations
appear to favor higher cutoff; the model with $10^{-1} M_\odot$ cutoff
predicts 5 to 7 microlensing events detectable by OGLE with a typical
time scale of 20 days.  These results hint that there are few brown
dwarfs in the Bulge.  A larger sample will enable more definitive
determination of the bulge mass function.

In summary, stars in the galactic bar are the major source of optical
depth for microlensing in the Baade's window fields monitored by the
OGLE program.  Using a self-consistent bar model that has been fit to
the DIRBE observations of the bulge surface brightness distribution
and to the observed stellar kinematics, we have computed the optical
depth towards Baade's window and the predicted event duration
distribution.  We find that the bar model provides a better fit to the
microlensing observations than an axisymmetric model for the Galaxy.
The optical depth of the bar model is consistent with the OGLE value.
The OGLE observed event duration distribution also favors models with
few brown dwarfs in the Bulge.

Future observations of stars in the bulge will test the hypothesis
that ordinary stars in the bar are the dominant microlenses.  When the
MACHO efficiency is quantified, the event duration distribution in
this experiment can also be compared to our theoretical predictions.
Since the lens probability is strongly peaked near 6.25 kpc in our
model, most of the lenses should be observable by Hubble Space
Telescope: for $A_I$=1 magnitude in Baade's window, 0.1 $M_\odot$
stars ($M_I \simeq $ 12) should have I$\simeq$ 27.  Eventually, the
detection of microlensing events at several fields can determine the
distribution of lenses and provide a definitive determination of the
nature of the dominant microlenses.

\acknowledgments

We would like to thank Bohdan Paczy\'nski for helpful comments and for
encouraging us to consider the implications of triaxiality for
microlensing and Kris Stanek for comments on a preliminary draft.  DNS
is partially supported by NSF grant AST 91-17388 and NASA grant ADP
NAG5-2693.  HSZ and RMR acknowledge support from Long-Term Space
Astrophysics grant NAGW-2479 to RMR.

\vfill
\eject
\begin{figure}[h]
\caption{shows the predicted transverse velocity dispersion
$\sigma_t$ (solid line) and the absolute value of transverse rotation
speed $V_t$ (dashed line) in km s$^{-1}$, and the arbitrarily
scaled probability of lens location (dotted line) as functions of
line-of-sight distance along Baade's window.  Note that the bar model
predicts that most of the lenses are at 6.25 kpc in the Bulge, where
$\sigma_t$ is also low.  The break in lens density near 4 kpc is due
to truncation.}
\end{figure}

\begin{figure}[h]
\caption{ shows event duration distributions of the OGLE data (histogram)
and models with the lower mass cutoffs at $10^{-1} M_\odot$ (solid
line) and $10^{-2} M_\odot$ (dotted line).  The upper panel shows the
fraction of events $f(<t_0)$ with duration shorter than $t_0$.  The
lower panel shows the logarithm of $P(t_0)$, the predicted rate of
microlensing events per $10^6$ bulge stars per year, as function of
time scale $t_0$.  The observations appear to favor the model with
higher cutoff.}
\end{figure}

\end{document}